\pdfoutput=1

\documentclass{iaus}
\usepackage{graphicx}
\usepackage{dcolumn}
\usepackage{bm}
\usepackage{natbib}

\newcommand{\Msun}{M$_{\odot}$}

\newcommand{\Mch}{M$_{\mathrm{Ch}}$}

\newcommand{\Tmax}{$T_\mathrm{max}$}
\newcommand{\rhoTmax}{$\rho(T_\mathrm{max})$}

\newcommand{\rhocd}{$\rho_\mathrm{d,c}$}
\newcommand{\rhoca}{$\rho_\mathrm{a,c}$}
\newcommand{\rhoc}{$\rho_\mathrm{c}$}

\newcommand{\rhocrat}{$\rho_\mathrm{d,c}/\rho_\mathrm{a,c}$}
\newcommand{\MencTmax}{$M_\mathrm{enc}(T_\mathrm{max})$}

\newcommand{\Tc}{$T_\mathrm{c}$}

\title[Properties of CO WD Merger Remnants] 
{Properties of Carbon-Oxygen White Dwarf Merger Remnants}

\author[Chenchong Zhu, Philip Chang, Marten van Kerkwijk and James Wadsley]   
{Chenchong Zhu$^1$, Philip Chang$^2$, Marten van Kerkwijk$^1$ \and James Wadsley$^3$}

\affiliation{$^1$Department of Astronomy \& Astrophysics, University of Toronto, \\ 50 St. George Street, Toronto, Ontario, Canada, M5S 3H4 \\[\affilskip]
$^2$Department of Physics, University of Wisconsin-Milwaukee, \\ 1900 E Kenwood Blvd., Milwaukee, Wisconsin 53211, USA \\[\affilskip]
$^3$Department of Physics \& Astronomy, ABB-241, McMaster University, \\ 1280 Main St. W, Hamilton, Ontario, Canada, L8S 4M1\\[\affilskip]
Correspondence email: {\tt cczhu@astro.utoronto.ca}}

\pubyear{2011}
\volume{281}  
\pagerange{0--0}
\jname{Binary Paths to Type Ia Supernovae Explosions}
\editors{R. Di Stefano \& M. Orio, eds.}
\begin{document}

\maketitle

\begin{abstract}
Recent studies have shown that for suitable initial conditions both super- and sub-Chandrasekhar mass carbon-oxygen white dwarf mergers produce explosions similar to observed SNe Ia.  The question remains, however, how much fine tuning is necessary to produce these conditions.  We performed a large set of SPH merger simulations, sweeping the possible parameter space.  We find trends for merger remnant properties, and discuss how our results affect the viability of our recently proposed sub-Chandrasekhar merger channel for SNe Ia.
\end{abstract}

\firstsection 

\section{Introduction}

The nature of type Ia supernova (SN Ia) progenitors, and their evolution leading up to the explosion, are still not yet well understood \citep{vkercj10}.  The most widely accepted scenario is that of a carbon-oxygen white dwarf (CO WD) in a binary that acquires mass through either accretion from a non-degenerate companion (the single-degenerate channel) or merger with another WD (the double-degenerate channel), until it approaches the Chandrasekhar mass, {\Mch} \citep{howe11}.  This scenario still has several unresolved issues, including discrepancies between the rates of SNe Ia and progenitors \citep{maoz08,ruitbf09}, the lack of sufficient numbers of supersoft sources for the single-degenerate (and potentially double-degenerate) channel \citep{dist10a,dist10b,gilfb10}, and the still-poorly understood nature of the deflagration-to-detonation transition, required for {\Mch} explosions to look like observed SNe Ia \citep{howe11}.  These issues suggest the utility of considering alternate SN Ia channels, for instance the proposed channels of \cite{pakm+10}, \cite{guil+10}, and, for this paper, \cite{vkercj10}.

\citeauthor{vkercj10} propose an SN Ia channel that involves the merger of two CO WDs whose total mass is below {\Mch}.  As shown in \cite{loreig09}, the merger between two equal mass 0.6 {\Msun} WDs results in a massive, uniformly rotating core with a differentially rotating, sub-Keplerian, fat disk.  While the merger is hottest at the centre of the core, it is not hot enough to ignite runaway carbon fusion.  The disk then accretes onto the core on a characteristic timescale of several hours.  The accretion compresses the core, driving up temperatures until the accretion and carbon fusion timescales are comparable.  The neutrino cooling timescale is six orders of magnitude longer, and therefore a carbon fusion runaway results.

This channel possesses a number of advantages.  If most merging CO WD binaries (regardless of total mass or mass ratio) are assumed to result in SNe Ia, it would account for the observed SN Ia rate \citep{vkercj10}.  Since lower-mass CO WDs generally come from lower-mass stars, it also explains the observed decrease of typical SN Ia luminosity with increasing host stellar population age.  \cite{sim+10} recently showed that sub-Chandrasekhar detonations can produce light curves and spectra very similar to observed SNe Ia, without resorting to a deflagration-to-detonation transition.  This channel stands in contrast to He-CO sub-{\Mch} merger channels, which involve the detonation of a helium layer around a sub-{\Mch} CO WD, producing light curves and spectra that do not resemble observed SNe Ia (\citealt{sim+10}; \citealt{howe11}; see e.g. \citealt{woosk11} on whether or not the effect of helium can be minimized).

In order to determine the viability of \citeauthor{vkercj10}'s analytical sketch, it must be determined which regions of the CO WD merger parameter space will lead to detonations, and whether, e.g., nearly equal mass WDs are required, and, if so, how close to unity the mass ratio must be.

We have run a series of 45 high-resolution smoothed-particle hydrodynamic (SPH) simulations of white dwarf mergers, exploring a broad range of parameter space to seek global trends in merger remnant properties as a means of interpolating between numerical results.  Our results can be used to create initial conditions for models of the further viscous and thermal evolution of the remnants (e.g. \citealt{shen+11}) to determine if any of them reach conditions necessary for carbon detonation.  They also have utility outside our own project, as they would allow anyone to estimate the properties of a CO WD merger remnant without having to simulate the merger themselves.

This report focuses on the implications of our results for the sub-{\Mch} merger channel.  Details regarding trends and a possible semi-analytical picture of CO WD mergers will be presented in a forthcoming publication.

\section{Code and Initial Conditions}

We used the SPH code Gasoline \citep{wadssq04}.  Gasoline uses the formulation for artificial viscosity in \cite{mona92}, augmented with a \cite{bals95} switch and a method that reduces the artificial viscosity coefficient $\alpha$ when no shocks are present.  We have modified Gasoline to use the Helmholtz equation of state (\citealt{cococubed}, details in \citealt{timms00}).  No nuclear reaction network was used, as previous simulations have shown that nuclear processing is not important to the hydrodynamic evolution of lower-mass mergers \citep{guerig04,loreig09}.  An \textit{in situ} detonation, however, may arise during mergers between nearly equal-mass CO WDs where the accretor has mass $\gtrsim 0.9$ {\Msun} \citep{pakm+10,pakm+11}.

Our input binaries are composed of $8 \times 10^4$ to $4 \times 10^{5}$ particles, depending on their mass.  White dwarfs were relaxed individually and given equilibrium temperatures of $\sim 10^7$ K.  Their initial separation is chosen such that the lower-mass binary fills its Roche lobe.  Since the synchronization timescale is expected to be much longer than the angular momentum loss timescale due to gravitational radiation, our binary is unsynchronized at the start of merger (\citealt{marsns04}, but see their discussion and \citealt{fulll11}).

\section{Results}

\begin{figure}
\centering
\includegraphics[angle=0,width=0.95\columnwidth]{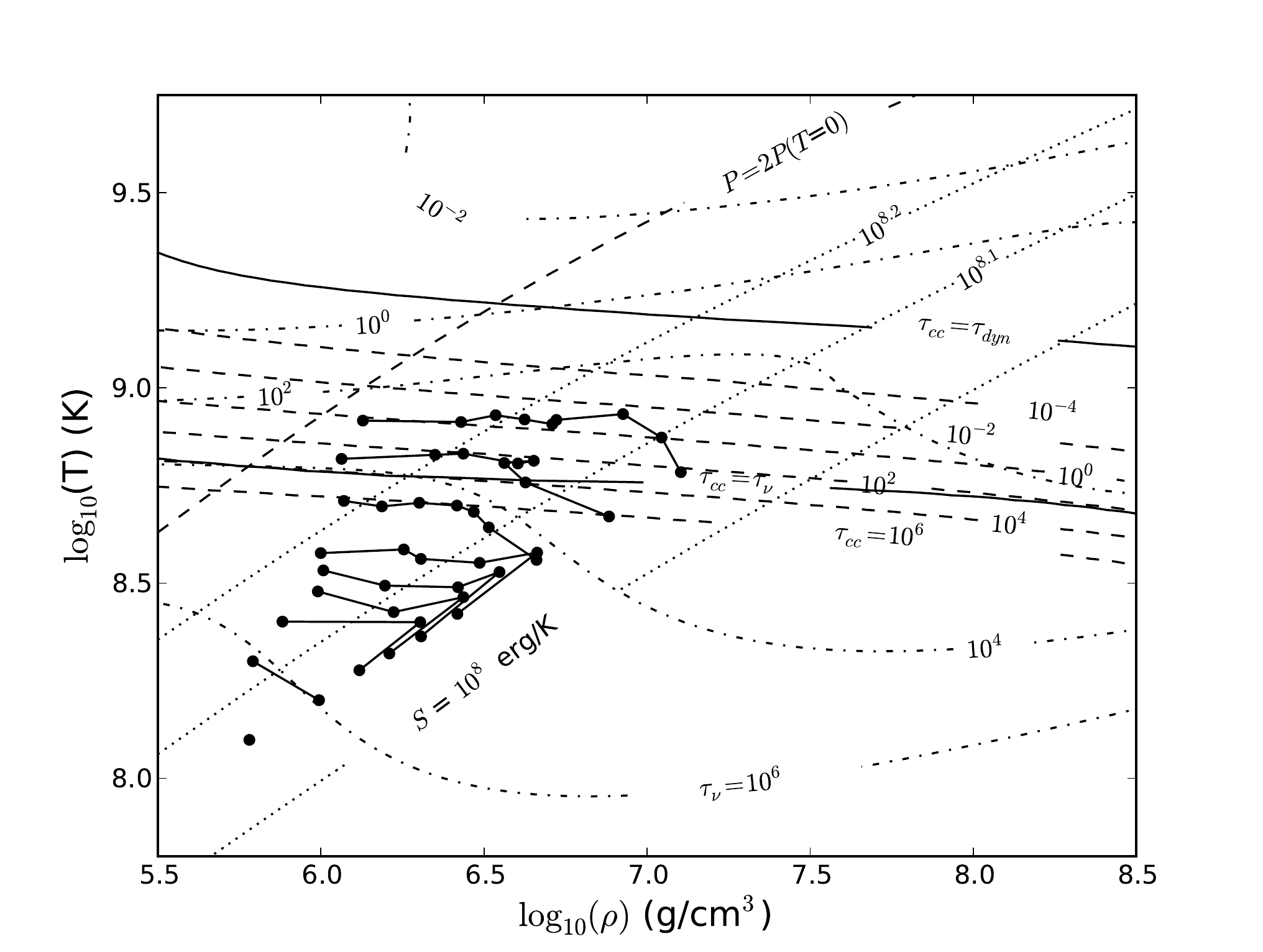}
\caption{Merger remnant maximum temperature and corresponding density.  Black points are merger remnants, and lines connect points with the same accretor mass (from bottom to top, 0.4, 0.5, 0.55, 0.6, 0.65, 0.7, 0.8, 0.9 and 1.0 {\Msun}).  Maximum temperatures of nearly equal mass mergers have been adjusted to account for mixing in their convectively unstable cores.  Also shown are contours of neutrino cooling timescale $\tau_\mathrm{\nu}$, carbon fusion timescale $\tau_\mathrm{cc}$ and entropy $S$.  The $\tau_\mathrm{\nu} = \tau_\mathrm{cc}$ and $\tau_\mathrm{dyn} = \tau_\mathrm{cc}$ lines denote where the carbon fusion timescale balances the neutrino cooling and dynamical timescales, respectively.  The $P = 2P(T$=$0)$ line marks the upper bound to where matter is degenerate.}
\label{willitexplode}
\end{figure}

The merger remnants for all our runs consist of a core and a sub-Keplerian disk.  For highly unequal mass mergers, almost the entire accretor forms 85 - 90 \% of the core, while $\sim70$\% of the donor forms the disk; for equal mass mergers, $\lesssim10$\% of donor and accretor go into the disk.  Other mergers fall on a continuum between these two extremes.  Some results important to the possibility of the remnant exploding are summarized below:

\begin{itemize}
	\item The \textbf{maximum temperature} {\Tmax} of the system is nearly independent of mass ratio, and is within 20\% of $aGM_\mathrm{a}m_\mathrm{H}/(R_\mathrm{a}k_B)$ (i.e. it scales with the gravitational potential of the accreting WD before merger), with $a=0.270$.  Most of the equal and nearly equal mass mergers have convectively unstable cores.  Artificially mixing these cores to make them isentropic decreases their maximum temperatures by 10-40\% (see Fig. \ref{willitexplode}).
	\item The \textbf{density at maximum temperature} {\rhoTmax} of the system is within 52\% of central density of the donor, {\rhocd}.
	\item The \textbf{central density of the remnant} {\rhoc} is within 67\% of {\rhoca}, the central density of the accretor.  The \textbf{central temperature} {\Tc} for unequal mass mergers is significantly lower than {\Tmax}.  Since our conditions are far from the pyconuclear regime, temperature is more important than density, and so {\Tmax} is more relevant to post merger evolution than {\Tc}, despite {\rhoc} being higher than {\rhoTmax}.
	\item The \textbf{enclosed mass inside the radius of highest temperature} {\MencTmax} is within 35\% of $M_\mathrm{a}$, the mass of the accretor, when {\rhocrat} $<$ 0.6, i.e. for unequal mass mergers the hottest point is near the core-envelope interface.  {\MencTmax} is zero when {\rhocrat} $>$ 0.6, i.e. the hottest point is at the centre for nearly equal mass mergers.  The cut-off at {\rhocrat} = 0.6 is sharp, suggesting that this ratio defines whether or not the merger is ``equal mass''.
\end{itemize}

\section{Implications for the Sub-{\Mch} Merger SN Ia Channel}

For Fig. \ref{willitexplode} we have replicated \citeauthor{vkercj10}'s Fig. 1, and added our simulated merger remnants.  Although several remnants are above the $\tau_\mathrm{cc} = \tau_\nu$ runaway line, none reach a carbon burning timescale shorter than the accretion timescale of several hours.  Accretion of the disk onto the core is therefore of most immediate consequence.  Accretion results in adiabatic compressional heating, and may increase density by up to an order of magnitude.  Transposing the points on Fig. \ref{willitexplode} one order of magnitude in density along a contour of constant entropy results in all remnants with accretor masses $\gtrsim 0.55$ {\Msun} reaching ignition ($\tau_\mathrm{cc} = \tau_\nu$).  Those remnants with small mass ratios $q \lesssim 0.5$ will likely not explode, since as the runaway progresses, they would lose degeneracy around the time the fusion timescale becomes equal to the degeneracy timescale ($\tau_\mathrm{dyn} = \tau_\mathrm{cc}$).  For all others, explosion seems, at least in principle, possible.

A more accurate depiction of post merger evolution would require taking into consideration the the full viscous evolution, angular momentum transport, heating and compression throughout the remnant (see \citealt{shen+11} for initial studies).  Likewise, the later, rapid nuclear runaway requires simulation to see if it can lead to detonation (Chang, White \& Van Kerkwijk, in prep.).  As noted earlier, merger remnants with their hottest point at the centre have {\rhocrat} $>$ 0.6 (corresponding to approximately $M_\mathrm{a} - M_\mathrm{d} \lesssim 0.08$ {\Msun}).  If this is required for an explosion, only a fraction of our remnants will explode.  It is still possible, however, that some systems will experience off-centre detonations as a result of compressional heating.  The heaviest of our equal-mass mergers may also explode as they merge, as described in \cite{pakm+10,pakm+11}.

Nevertheless, Fig. \ref{willitexplode} suggests that the sub-{\Mch} merger channel for SNe Ia merits further investigation.

\end{document}